\documentclass[runningheads]{llncs}
\usepackage{graphicx}
\usepackage{algorithm}
\usepackage{algorithmic}
\usepackage{amsmath}
\usepackage{float}
\usepackage{color, colortbl}

\usepackage{amsmath,amssymb,amsfonts}
\usepackage{algorithmic}
\usepackage{graphicx}
\usepackage{textcomp}
\usepackage[caption = false]{subfig}
\begin{document}
\title{An Overview of Blockchain and 5G Networks}

\author{Hajar Moudoud\inst{*,1,3} \and
Soumaya Cherkaoui\inst{1} \and
Lyes Khoukhi\inst{2}}
%

%
\institute{Université de Sherbrooke, Canada\\
\email{{hajar.moudoud, soumaya.cherkaoui}@usherbrooke.ca}\and
Normandie University, France\\
\email{lyes.khoukhi@ensicaen.fr} \and
University of Technology of Troyes, France\\
}
\maketitle     

\abstract{The 5G wireless networks are potentially revolutionizing future technologies. The 5G technologies are expected to foresee demands of diverse vertical applications with diverse requirements including high traffic volume, massive connectivity, high quality of service, and low latency. To fulfill such requirements in 5G and beyond, new emerging technologies such as SDN, NFV, MEC, and CC are being deployed. However, these technologies raise several issues regarding transparency, decentralization, and reliability. Furthermore, 5G networks are expected to connect many heterogeneous devices and machines which will raise several security concerns regarding users' confidentiality, data privacy, and trustworthiness. To work seamlessly and securely in such scenarios, future 5G networks need to deploy smarter and more efficient security functions. Motivated by the aforementioned issues, blockchain was proposed by researchers to overcome 5G issues because of its capacities to ensure transparency, data reliability, trustworthiness, immutability in a distributed environment. Indeed, blockchain has gained momentum as a novel technology that gives rise to a plethora of new decentralized technologies. In this chapter, we discuss the integration of the blockchain with 5G networks and beyond. We then present how blockchain applications in 5G networks and beyond could facilitate enabling various services at the edge and the core.}


\section{Introduction}
\label{sec:1}
Fifth-generation (5G) networks and beyond are expected to enable a wide range of applications from the industrial internet of things, virtual reality, autonomous driving to real-time gaming \cite{s1}. will enable them to benefit from massive machine-like communications, improve mobile broadband, and provide ultra-reliable, low-latency communications. The 5G roadmap aims to provide users with up to 10 Gbps of data throughput, 1000x network capacity, 10 Tbps per square kilometer, and 1 millisecond latency \cite{r3}. To meet these requirements, several underlying wireless technologies have been proposed, such as cloud computing (CC), multi-access edge computing (MEC), software-defined network (SDN), and network function virtualization (NFV). In addition to these new technologies, 5G will incorporate new radio access techniques, including massive MIMO, millimeter wave and D2D connectivity, which will also enable 5G cellular networks.

The 5G networks will support new technologies to deliver new service delivery models. These technologies, however, will cause several challenges regarding users’ security, data privacy, and integrity. Unlike centralized cellular networks ($i.e.,$ 3G, 4G, etc.), 5G networks are expected to distribute and decentralize services that emphasize security challenges \cite{r27}. Specifically, the security management in 5G networks is complex because it operates in a flexible and dynamic environment where a massive number of devices are connected. 

For secure, transparent, and immutable 5G networks, there is a need to ensure security while exploiting new technologies such as Distributed Ledger Technologies (DLT), Artificial Intelligence (AI), and Machine Learning (ML). Among the existing technologies, blockchain is the most promising one that has the potential to revolutionize the way services are offered 5G networks \cite{r4} \cite{bb}. Blockchain technology is a distributed, immutable, and secure ledger that ensures trust among unreliable entities without relying on a centralized third party. Blockchain has the capacity to be merged with the 5G to provide reliable resource sharing, secure storage, smart authentication, and security management \cite{r5}. Consequently, blockchain with its inherent features will provide massive communication in a distributed environment while ensuring high security, data privacy, and reliability.

Currently, one of the challenging aspects of 5G networks is the need to assure transparency, immutability, and decentralization for its large number of users and services. Blockchain technology with its inherent properties will enable secure and trustworthy data transactions in a P2P manner for various services/users in the 5G network \cite{r6} \cite{r23}. Therefore, blockchain integration with 5G networks will result in self-managing, self-securing, and self-maintaining networks without the need for a central authority. In fact, 5G is expected to provide a connection for a large number of devices with resources and services (network slicing). In this regard, distributed blockchain with its security features will enable a new generation of massive communication by providing transparent provisioning between devices.

Besides the benefits of blockchain integration with 5G networks, 5G will enable a wide range of blockchain-based applications such as autonomy's resource, automated management, and reliable content base storage \cite{r7} \cite{r22}. For instance, blockchain can be leveraged by cellular service providers (CSPs) to enable future services for mobile industries. In order to automatically handle transactions in CSPs, blockchain features such as smart contracts which are self-executing code will enable a secure and automated sharing of infrastructures. Furthermore, the blockchain and ML empowered intelligent 5G beyond network will enable a new stack of technologies that will empower self-aggregating communication and intelligent resource management beneficial for 5G networks.

\section{Background}
\label{sec:Background}

In this section, firstly we give a brief background on the blockchain and its features. Secondly, we give a brief description of blockchain features.
\subsection{Blockchain}
\label{subsec:Blockchain}
Blockchain is a transparent technology, that establishes trust among unreliable entities \cite{r10} \cite{r22}. This decentralized technology ensures secure information transmission without the interference of a third authority. Blockchain is a distributed ledger that keeps a continuously growing set of data records called blocks. Each block contains a collection of transactions committed by members of the blockchain. A blockchain transaction is a piece of information meant to be stored in a safe database \cite{r11}\cite{r21}. For instance, a transaction could contain technical metadata of a mobile ($i.e.$, size, type, or timestamp). The blockchain is an immutable ledger that permits transaction verification by members who could be dishonest. To add a new block to the blockchain, all members of the chain should reach an agreement; this is referred to as a consensus. Once a consensus is established among all members, the new block is validated then added to the chain.

Blockchain technology has many characteristics, such as being decentralized, immutable, and having no single point of failure (SPOF). Blockchain can operate in a decentralized environment, each member of the chain has an integral copy of the ledger, meaning, data is stored in a peer-to-peer environment \cite{r12}. This redundancy of information guarantees data non-repudiation, making it difficult to cause major disruption. Blockchain is designed to be immutable; once a block is added to the chain, any changes inside the block will be extremely difficult. The blockchain enables several technologies like the hash function, the digital signature, and the timestamp. If a malicious user wants to change a block, this will cause the hash to change also, meaning he needs to reach a new agreement for this block and other blocks following it. Due to the distributed and shared nature of the blockchain, the ledger cannot be controlled by a centralized entity, meaning, it has no SPOF.
 \subsubsection{Blockchain Taxonomy}
 \label{subsec:Blockchain Taxonomy}
A blockchain is composed of a family of blocks linked together by a hash. The hash function is used to map data stored in a transaction and generates a unique fingerprint \cite{r15}. The hash does not only depend on the new transaction but on the previous transactions also. A new transaction is broadcast to miners of the chain and waits to be confirmed. To verify a transaction, miners need a digital signature to certify the authenticity and integrity of the transaction. The blockchain uses the Elliptic Curve Digital Signature Algorithm to sign a transaction. Once the blockchain miners approve a transaction, it is written into a block. A block is added to the chain when the consensus is established, and the block has reached a certain number of verified transactions. Each block refers to the previous block and together form the blockchain. Figure 1 presents how blocks are linked together.
\subsubsection{Blockchain Platform Types}
\label{subsec:Blockchain Platform Types}
Current blockchain networks can be classified into three types: public, private, and consortium Blockchain \cite{r13}. A public blockchain is permissionless where all members can join the network. It is publicly open for members to read, write, or validate a transaction without the approval of third parties. In a private blockchain, only members can participate in the network, meaning, it is centralized. The owner of the consortium blockchain, only a group of authorization members can validate transactions \cite{r16}. These members can be chosen.
in advance.
\subsubsection{Blockchain Consensus}
\label{subsec:Blockchain Consensus}
The consensus is a protocol that establishes the core of the blockchain and decides how the agreement should be reached among miners to append a new block to the chain \cite{r14}. Blockchain consensus can be defined as an algorithm that helps the distributed network to make decisions, while others describe the consensus algorithm as the mechanism that brings about all nodes to take a decision about the same transaction either to add it to the chain or reject it. According to others making a consensus among unreliable miners is a transformation of the Byzantine generals. There are several types of blockchain consensus. The most common consensus algorithms are Proof of Work (PoW), Proof of Stake (PoS), Proof of Activity (PoA), and Practical Byzantine Fault Tolerance (PBFT).
\begin{figure*}[t]
	\centering
	\includegraphics[width=\linewidth]{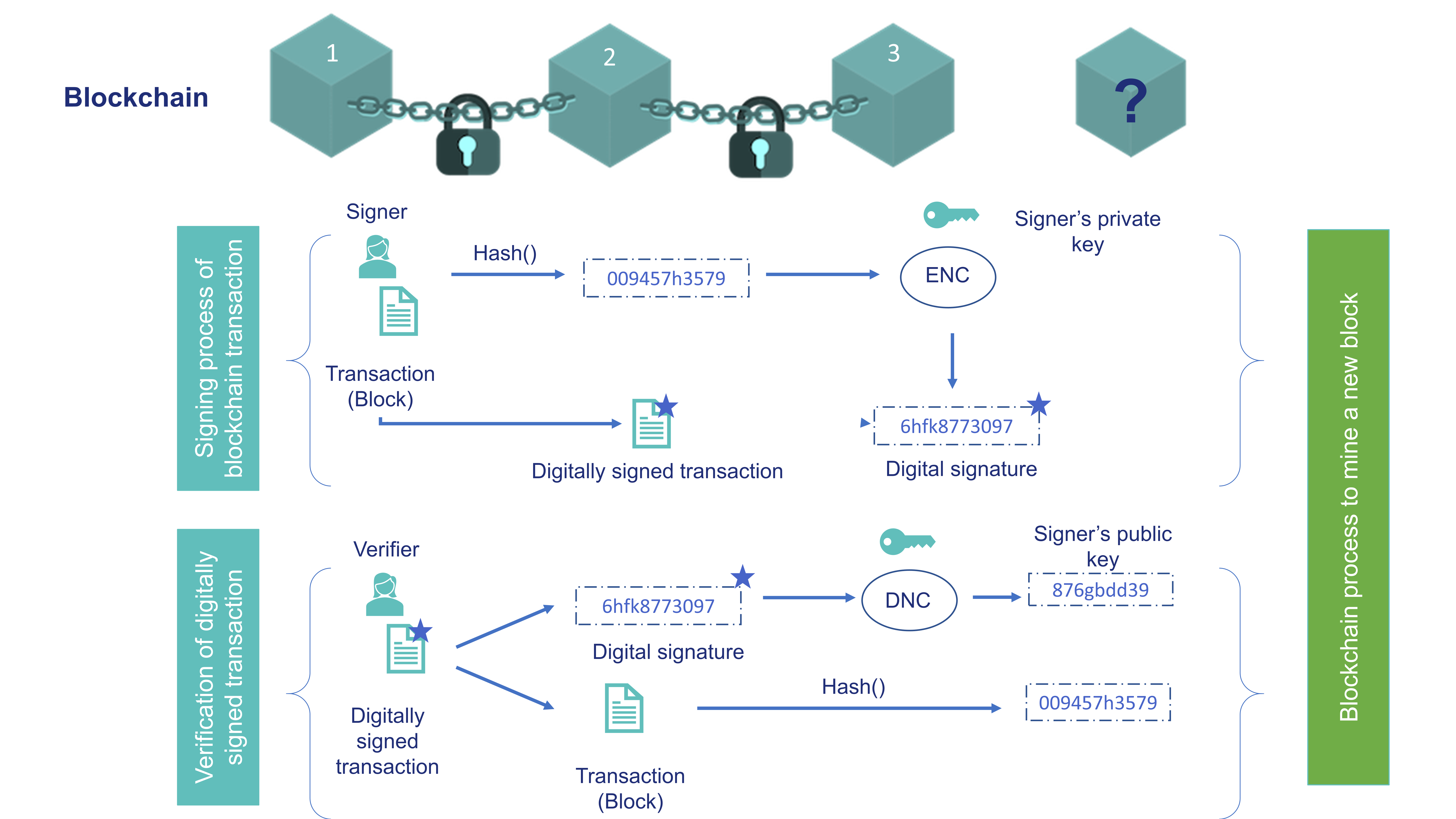}
	\caption{Blockchain workflow.}
	\label{fig:archi}

\end{figure*}
\subsubsection{Blockchain Smart contract}
\label{subsec:Blockchain Smart contract}
Blockchain smart contracts are self-executing programs non-modifiable implemented inside the chain which is intended to automatically execute actions to the terms of a contract
\cite{r17}. Smart contracts are implemented permanently inside the blockchain to execute a code once some conditions are verified. The most used platform to run smart contracts is the Ethereum blockchain. These smart contracts are coded using a coding language called Solidity. For example, blockchain smart contracts can be implemented to help to manage a system or to establish access control inside a network \cite{icc6}. The smart contracts can run dynamically in the network.
\subsubsection{Blockchain Sharding}
\label{subsec:Blockchain Sharding}
The sharding consists of splitting a large collection across several servers, enabling the distributed management for the collection, thereby improving the scalability \cite{r19}. Blockchain sharding refers to the artificial division of the workload of the transaction procession into a single shard, this way one single transaction can be validated and stored by many members working in parallel. Yet, applying sharding to the blockchain has several challenges, such as weak security, add throughput, and increased risk for data loss.
\subsubsection{Blockchain Oracle}
\label{subsec:Blockchain Oracle}
Blockchain cannot access external data of the network. This is where the blockchain oracle intervenes. It is a service provider (trusted third party) that verifies the data authenticity and attests to facts in an effort to bring outside world data into the chain.
\section{5G Networks and Beyond: An Overview}
In this section, we present an overview of the 5G networks.
\subsection{Software-Defined Networking (SDN)}
SDN enables external control of data away from network hardware to software referred to as the controller. The controller manages packet flow to provide intelligent networks. With the controller, users will be able to manage network equipment using software, and thus introducing new services. An intelligent 5G network will enable new operations and offer new services on demand while ensuring efficiency. 5G SDN will be able to control and orchestrate services in a seamless and efficient manner. SDN architecture will provide high flexibility to 5G networks permitting it to be perfect for the dynamic bandwidth nature of 5G.

\subsection{Network Function Virtualization (NFV)}
NFV refers to the replacement of hardware infrastructure by virtualization software and for different network functions ($e.g.,$ VPN, load balancers, firewall, routers, switches). NFV decouples the network functions from physical infrastructure and permits it to run virtually on a cloud infrastructure. The key objective of NFV is to transform the way networks are built and services are delivered. In 5G, NFV will enable the distributed cloud, helping leverage scalability, dynamicity, and flexibility.
The SDN and NFV are complementary to each other and both helping network abstraction and virtualization. The difference between NFV and SDN is that NFV isolates network control functions from network forwarding functions, whereas NFV isolates the network functions from physical infrastructure to the cloud.

\subsection{Network Slicing}
Network slicing enables multiple networks to work virtually over one physical network infrastructure. Each network slice is isolated from the physical network to meet the requirements requested of an application. Integrating NFV with network slicing will allow multiple applications/services to be deployed for users. Consequently, applications or services running over a network slice can display a high quality of experience (QoE) and high-quality services (QoS) beneficial for users.
\subsection{Multi-Access Edge Computing (MEC)}
MEC reduces network congestion and thus achieves a faster response. It enables computing capabilities of the cloud to the network edge which allows data to be processed near to the devices and thus users. Furthermore, by enabling the computing locally, MEC reduces significantly the energy need to process data and storage space. MEC enabled services, applications, and operations to be closer to the users which enhances the QoS for users; it enables network collaboration which improves the quality of experience.
\subsection{Device to Device (D2D)}
In current cellular networks, base stations are responsible for establishing communications between two devices. These communications are going through base stations even if the two devices are in the same ranges. Consequently, the spectral efficiency is low because of the delay added, which is unsuitable for real-time applications. To solve this issue, the 5G networks proposed using D2D communications to allow two devices close to each other to communicate using a direct link. This concept creates multi-hop relays among several devices which increase the data rate and improve QoS. Furthermore, D2D connectivity will assist 5G networks to be more malleable in terms of offloading and energy-efficient because it eliminates unwanted traffic from the core network.
\subsection{Cloud Computing (CC)}
CC was proposed to achieve the ever-growing requirements of 5G networks, such as resource orchestrations, data storage, and mobile sensing. CC enables resource offloading by virtualizing physical infrastructure to dynamically provide 5G services with their requests. Cloud computing includes two tiers: 1) Infrastructure Providers which manages the physical infrastructures (InPs), and 2) Service Providers (SPs) which provides services to network users.

\section{Blockchain for 5G}
In this section, we discuss the benefits at various levels of blockchain integration with 5G networks. Then, we present a taxonomy of the opportunities brought by blockchain applications on 5G.
\begin{figure*}[t]
	\centering
	\includegraphics[width=\linewidth]{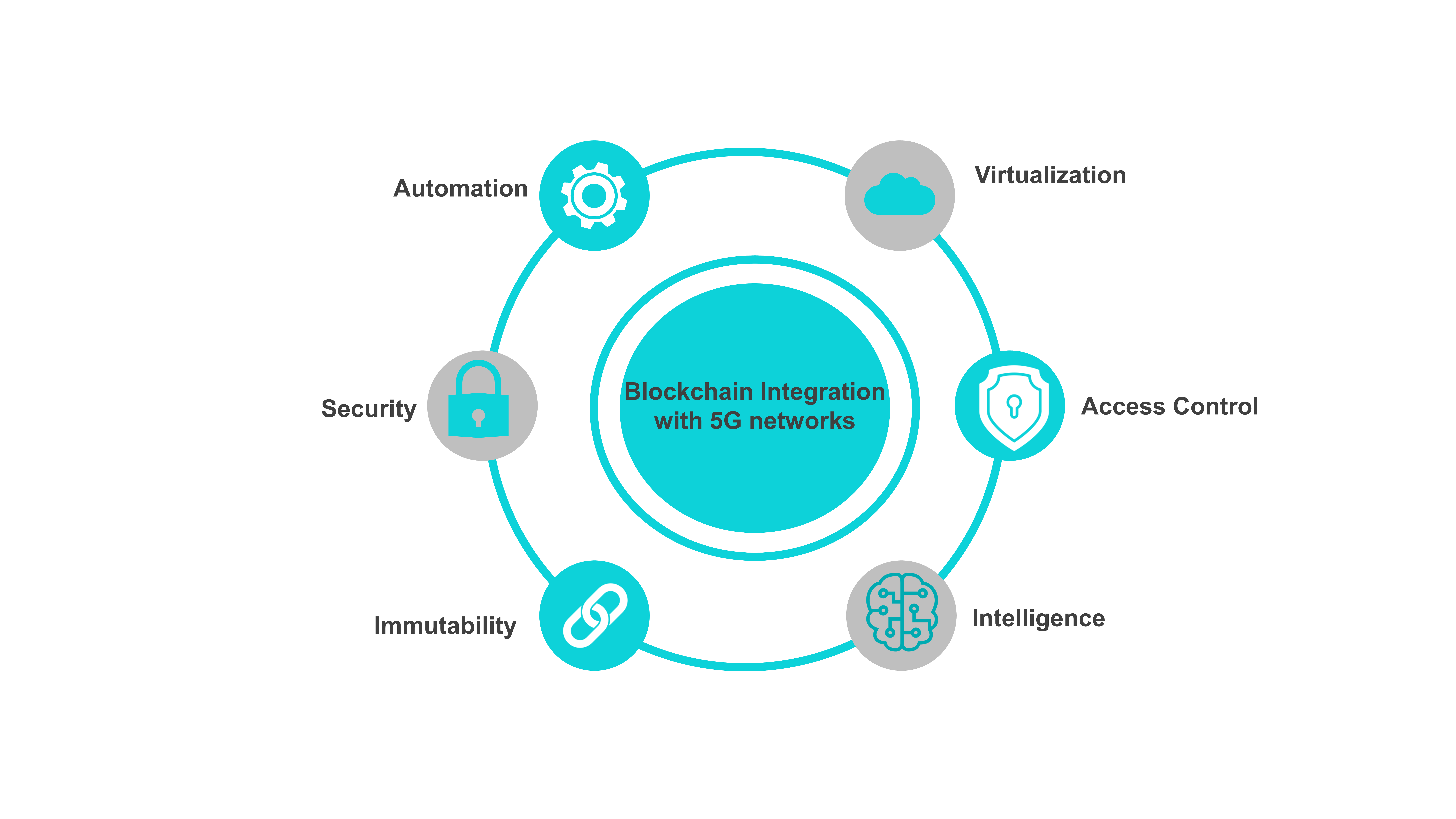}
	\caption{Blockchain integration with 5G opportunities tree.}
	\label{fig:archi}

\end{figure*}

\subsection{Blockchain Integration with 5G Networks}

5G and blockchain are two prominent technologies that could reshape the future of technologies and the telecommunication sector. On one hand, blockchain can provide a distributed environment to secure 5G services, operations, and data. Blockchain techniques can assist 5G to enable features such as decentralization, immutability, and transparency. On the other hand, 5G networks are highly distributed and require enabling new technologies such as NFV, SDN, D2D, MEC, and CC. These technologies are complex to orchestrate and manage. Furthermore, 5G networks share resources, services, and operations among several stakeholders that could be dishonest. Consequently, blockchain will enable future 5G networks with a high level of security, coordination, and manageability required among 5G users. All these new features will present challenges that need to be solved for the safe deployment of 5G networks.

The reason behind the integration of blockchain with 5G networks comes for the most part from the prominent features of blockchain that could solve the challenges in 5G networks in terms of security, privacy, management, and transparency. Because 5G involves several stakeholders, it is difficult to leverage trust while preserving their security. With the use of blockchain technology, 5G stakeholders are secured without the need of trusting each other. Blockchain consensus establishes trust among unreliable entities without the use of a trusted third party. Furthermore, shared services and operations are secured using the blockchain immutable ledger. Indeed, once data is stored inside the blockchain ledger it cannot be altered or falsified, blockchain uses cryptographic signature and hash function to secure data. Moreover, blockchain smart contracts can be used to automate the management of 5G resources and services. The entire process is transparent, reliable, and decentralized. Fig. 3 presents a conceptual diagram for the integration of blockchain with 5G networks. Next, we present the opportunities brought by the integration of blockchain with 5G.
\begin{figure*}[t]
\centering
\includegraphics[width=\linewidth]{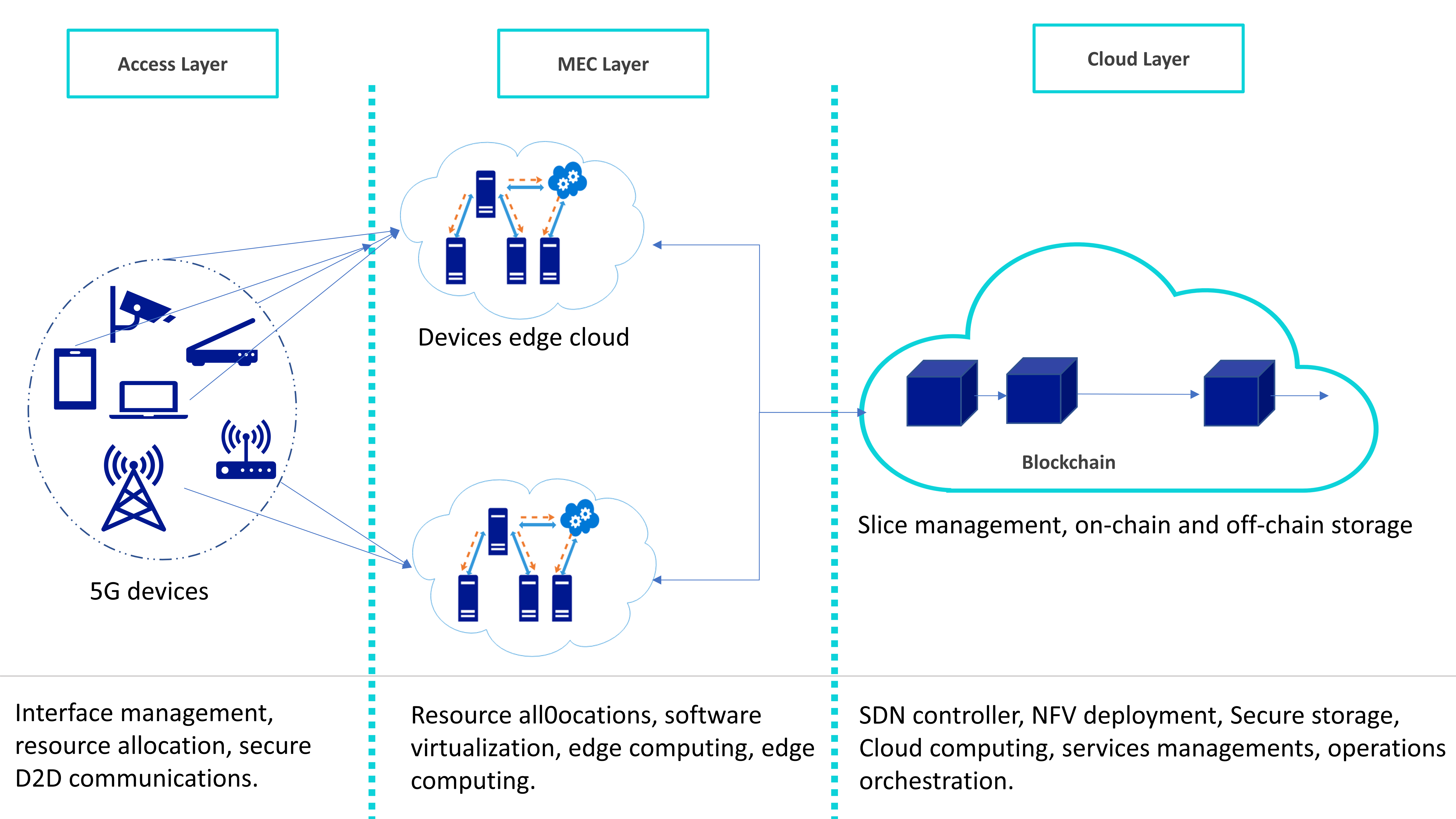}
\caption{Blockchain merging with 5G networks.}
\label{fig:archi}

\end{figure*}
\subsection{Opportunities brought by blockchain integration with 5G networks}
Blockchain interesting features will provide a new set of solutions to improve 5G networks in terms of security, transparency, immutability, privacy, and interoperability. Consequently, 5G networks should take advantage of blockchain to improve their flexibility, manageability, and security of network services and operations. Next, we present the opportunities that blockchain technology can conduct to 5G networks with regard to security improvements and performance enhancements. 
\subsubsection{security improvements}
5G networks face several security challenges, from data privacy to authentication vulnerability such as frequent authentication applied to ultra-dense networks \cite{r18}. These challenges expose 5G networks to attacks and increase their impact. Especially, when services and operations are shared among several stakeholders when one service is under attack, all the stakeholders taking part are exposed to attacks. Fortunately, blockchain features will improve 5G networks in terms of security, data privacy, and access control \cite{z1}. For example, blockchain Byzantine Fault Tolerance (BFT) consensus can help 5G networks to achieve trust in a distributed network even when some of the stakeholders in the network respond with incorrect information. Furthermore, blockchain can decentralize network management without the need for a third-party authority. For instance, the use of blockchain-based cloud computing can enable the decentralization of MEC 5G networks which take out the control from the core network, provide decentralize management, eliminates SPF issues, and improves trust in the network. In addition, blockchain can help to secure D2D communication by building a centralized peer-to-peer blockchain network, which considers each device as a blockchain miner that holds a copy of the ledger, verifies the authenticity of a transaction, and monitor transactions for better system reliability.

5G networks, with high connectivity and low latency, comes to support distributed new services and operations, which need decentralized management, limited access control, secure data sharing, and authentication services. Blockchain with smart contracts can establish a decentralized access control both for stakeholders and services. Because blockchain uses computing power to establish trust, data sharing, resource allocation, and spectrum sharing can be strongly secured against attacks such as False Data Injection or Modification Data Attacks. Indeed, several researchers have proven the efficiency of blockchain technology to secure 5G networks in terms of better access control using blockchain contracts, secure data sharing using hash functions, and cryptographic signatures. Furthermore, blockchain can host several technologies like Artificial Intelligence (AI) that could work on their own. Applied on 5G networks, blockchain, and AI can be able to operate in an independent manner to ensure better network orchestration and service management. All without the need of a trusted third authority.

With the support of blockchain features, 5G networks can fully benefit from better access control to services, secure data storage, and improved network management using blockchain smart while ensuring transparency and privacy. Furthermore, the strong immutability of blockchain ledger can provide a high degree of security for 5G users especially when sensitive data are being shared at a high speed among several 5G stakeholders. Furthermore, blockchain smart contracts can provide efficient access control mechanisms and authentication solutions that could help improving authentication issues in 5G networks. Blockchain smart contracts can implement automated access rules that will help authenticate 5G stakeholders without relying on external authority, can limit malicious behaviors in the network, and can detect threats without leaking stakeholders’ information. Besides, because blockchain uses hash functions, data is signed making it difficult to corrupt or falsified. Blockchain is capable of proving high data protection when sharing among untrustworthy 5G stakeholders while ensuring transparency.
\subsubsection{performance enhancements}

Blockchain technology can improve the performance of 5G systems. First, blockchain miners can verify data access and credentials which helps 5G networks protecting data. Second, blockchain smart contracts can manage automatically 5G services with a low latency which may lead to reduce the management costs. Finally, the use of a decentralized blockchain can provide a flexible, efficient, and secure data delivery system suitable with the 5G complex environment. For instance, blockchain can leverage a peer-to-peer network to secure communication among all 5G stakeholders and ensure trust among participants using appropriate consensus algorithms. Overall, blockchain integration with 5G networks can reduce communication costs, latency, and provide a secure platform for data exchange for all stakeholders, all of which improves the overall system performance.

Interference management is a known factor that influences the performance of wireless mobile networks. Because of the dense deployment in 5G networks and cellular interference, interference management will be further complex and hard to overcome. Further, collaborative communications will play an important part in 5G networks, managing a huge number of resources, services, and stakeholders while ensuring a high quality of services will even be greater. In this context, blockchain can solve some of these issues by decentralized interference administration. For example, blockchain can implement a cooperative interference management algorithm in blockchain smart contracts to ensure mutual trust and coordination protocols of interference.

\section{A Scalable and Secure Blockchain Suitable for 5G}
We solely reviewed the integration of blockchain with 5G networks. In this section, we present our vision of merging blockchain with 5G. We provide an overview of the proposed architecture named Block5G.
\subsection{A Scalable and Secure Blockchain Architecture Suitable for 5G}
Following our presentation of blockchain opportunities that blockchain can bring to 5G networks, in this section, we present our vision of a blockchain integration with 5G networks. 5G network needs a secure, scalable, and non-computing intensive solution that meets the requirement of low-latency, high communication rate provided by the network. In this context, we present a scalable and secure blockchain architecture that uses the sharding concept, and blockchain oracles suitable with 5G requirements named Block5G.

\subsection{Architecture}
As illustrated in Fig.4, we consider data exchange in a 5G network where a large number of unreliable users are connected through a D2D communication. We propose a scalable, secure, and trustworthy blockchain architecture composed of three layers, as follows:(1) the access layer which includes 5G devices that send and receive data; (2) the edge layer which is responsible for forwarding packets and verifying their validity using a blockchain consensus and blockchain oracles; and (3) cloud layer which is responsible for storing the data and scaling the blockchain while keeping security guarantees.
\subsubsection{Shared Blockchain}
The scalability has been a core problem in the integration of blockchain within 5G. This is because the 5G includes a large number of devices and users that communicate at a high rate and thus generate big data. To enable horizontal scalability, blockchain sharding was proposed, it consists of partitioning each transaction into several shards and processing it independently. In this paper, we propose building clusters with multiple nodes ($i.e.$miners) to process each shard in parallel.
\begin{figure}[t]
	\centering
	\includegraphics[width=\linewidth]{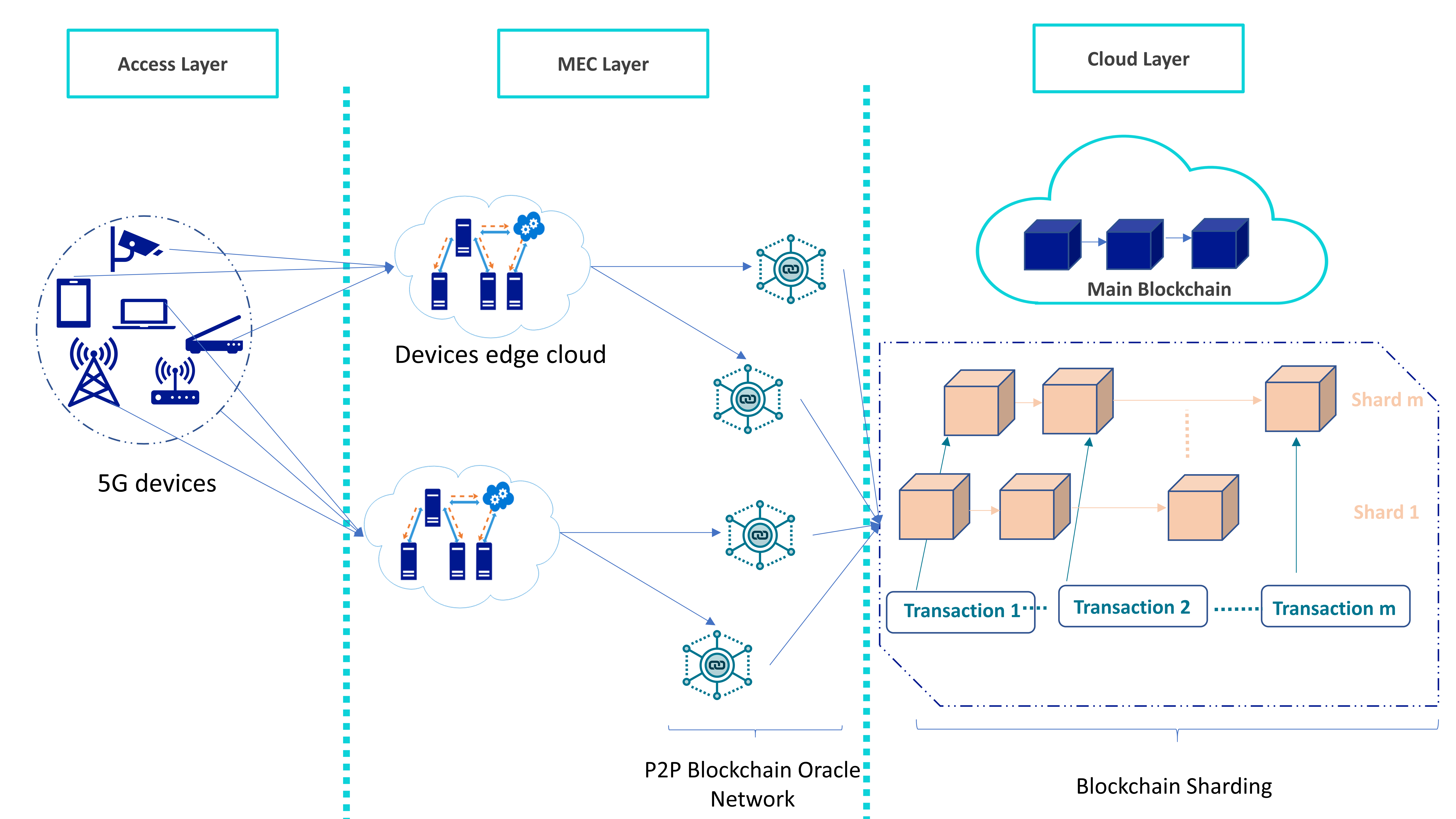}
	\caption{Scalable, trustworthy, and secure blockchain suitable with 5G.}
	\label{fig:archi}
\end{figure}
\begin{figure}[t]
	\centering
	\includegraphics[width=\linewidth]{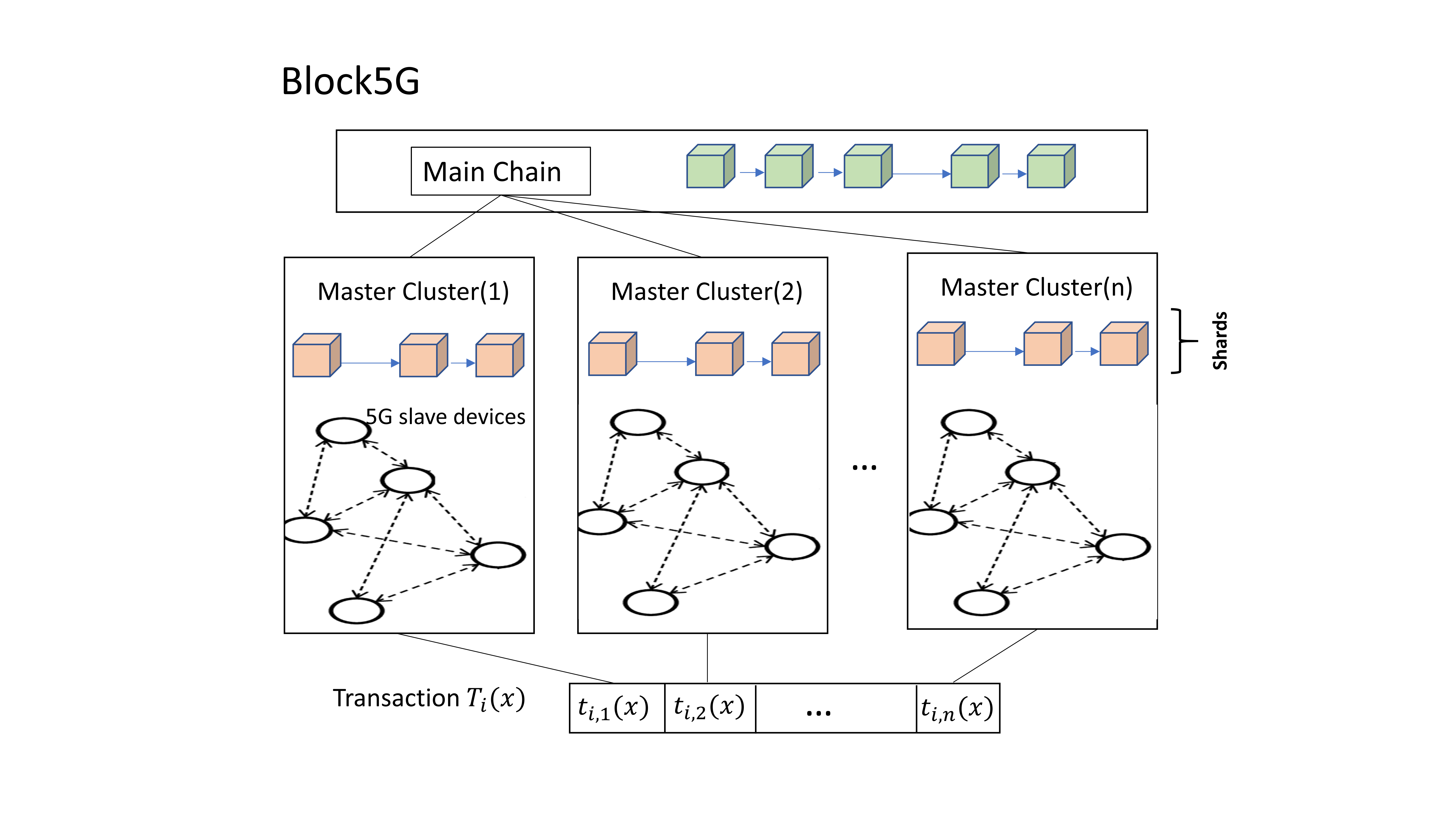}
	\caption{Scalable and Trustworthy Blockchain sharding in 5G context.}
	\label{fig:archi}
\end{figure}
Fig.5 shows the 3 main components of the Block5G sharding concept, as follows: (1) the main chain; (2) the master clusters'; and (3) slave devices. The master cluster contains $n$ slave devices that could be either honest or dishonest. Let $T_i(x)$ represents the $i_{th}$ transaction in a block. To enable the sharding, each transaction is divided into $n$ disjoint portions. All master clusters are responsible for verifying a portion of the transaction $t_{i,j}(x)$, where $i$ is the transaction number on a block and $j$ is the cluster number. All slave devices could verify a transaction portion, outputs 0 or 1 indicates invalid and valid transaction, respectively. Let $n$ denote the total number of slave devices in a master cluster $j$ that can tolerate up to $t \leqslant \frac{n}{3}$ dishonest devices. We define a transaction verification process as follows:
\begin{equation}
T_{i}(x)=  \frac{\sum_{k=0}^{m} t_{i,j}(x)}{m} 
\end{equation}
Where $m$ is the total number of miners that have participated in the verification process.

The transaction verification process has two possible outcomes: valid and invalid, each of which carries its own reward structure. As compensation for their efforts, the clusters are awarded ($R$) whenever they verify a transaction correctly. In the case of correct verification, a cluster reward is as follows.
\begin{equation}
R(T_{i,j}) =\frac{R(t_{i,j}(x))}{R(T_{i,Total}(x))} 
\end{equation}
Where, $R(t_{i,j}(x))$ and $R(T_{i,Total}(x))$ represents, respectively, the reward for a valid transaction portion and total reward for a valid transaction.
Note that the cluster trust value is only used to determine rewards and penalties, and does not necessarily correspond to the cluster output. Consequently, we enable secure sharding, prevent data loss and incentive clusters to behave honestly. 
\subsubsection{Peer-To-Peer Oracle Network}
We propose using a Peer-To-Peer (P2P) oracle network to verify the data queries and authenticate its source. Blockchain cannot access external data of the network. This is where the blockchain oracle interferes, it is a service provider (trusted third party) that verifies the data authenticity. However, trusting a single third party may lead to providing corrupt or inaccurate data. To this end, we propose using a P2P oracle network that ensures the truth value of 5G data. We assume each received data sent from a 5G device could be either valid $V$ or false $F$. There is $m$ oracle in the P2P network, only $n$ oracle verify the data. For each oracle $o$ $\in$ [n,m] has a $q$ probability that data $d$ is correct about a given proposition.

\begin{equation}
O_i(d)=
\begin{cases}
q \quad  \text{when data is valid}\\
1-q \quad \text{when data is false} 
\end{cases}
\end{equation} 
The value of $o_i(d)$ is independent of $o_j(d)$ for all $i \neq j$. In other words, each oracle's trust values are independent of other oracles in the network trust values. Furthermore, oracles need to place a deposit to participate in a random-chosen verification process. First, the oracle submits its verification probability $q$ $\in$ [0,1] about a data $d$. Then, this probability is applied to a trust weight $W$ $\in$ [0,1] that is, informally, the parameter within the network that verifies the input data within the network's hidden layers. Formally:
\begin{equation}
W_{i}=\frac{\alpha_i}{\beta_i}
\end{equation}
where $\alpha_i$ is the sum of corrected verification performed and $\beta_i$ is the total number of verifications performed. 
\begin{equation}
V(d)=\sum_{i=0}^{n}o_i(d)\times W_{i}
\end{equation}
Once a data verification process $V(d)$ has accumulated sufficient verifiability during a maximum of a period of time $\delta(t)$, it is decided. This period of time is a fixed value decided by network operators. The verification process has three possible outcomes:1) valid ($V$), if $V(d)$ value to be strictly positive, 2) false ($f$), if $V(d)$ value to be strictly negative, and 3) undefined ($U$) if $V(d)$ value is equal to zero. In this last case, the data verifiability process is only assigned to the oracle with the highest trust weight. 

Broadly speaking, oracles are rewarded when they participate in a verification process and their verification probability matches it. Conversely, those who gave incorrect verification are penalized. In the case of undetermined outcomes, oracles receive no rewards or penalties. As argued in the paper, the proposed data verification process incentives the oracles to behave honestly on the validity of data.
\subsection{Design Components}
The transaction verification procedure of our Block5G of the following (1) Initialization; and (2) Reward. This procedure starts with the initialization and then proceeds in periods to the rewarding phase. We now explain each component in more detail. 
\subsubsection{Initialization} The initial set of participants ($i.e.$, nodes) are invited to provide a deposit ($stake$) to validate or endorse a transaction. That is, a node is given the chance to verify a transaction chosen uniformly at random from the unverified transaction pool. The deposit is placed before the verification process. Because the nodes are grouped in clusters, the outcome of the transaction verification reward is a function of the sum of the total transaction verification reward weighted by the cluster value in the cluster and the deposits. 
\subsubsection{Reward}

\begin{figure}[t]
	\centering
	\includegraphics[width=\linewidth]{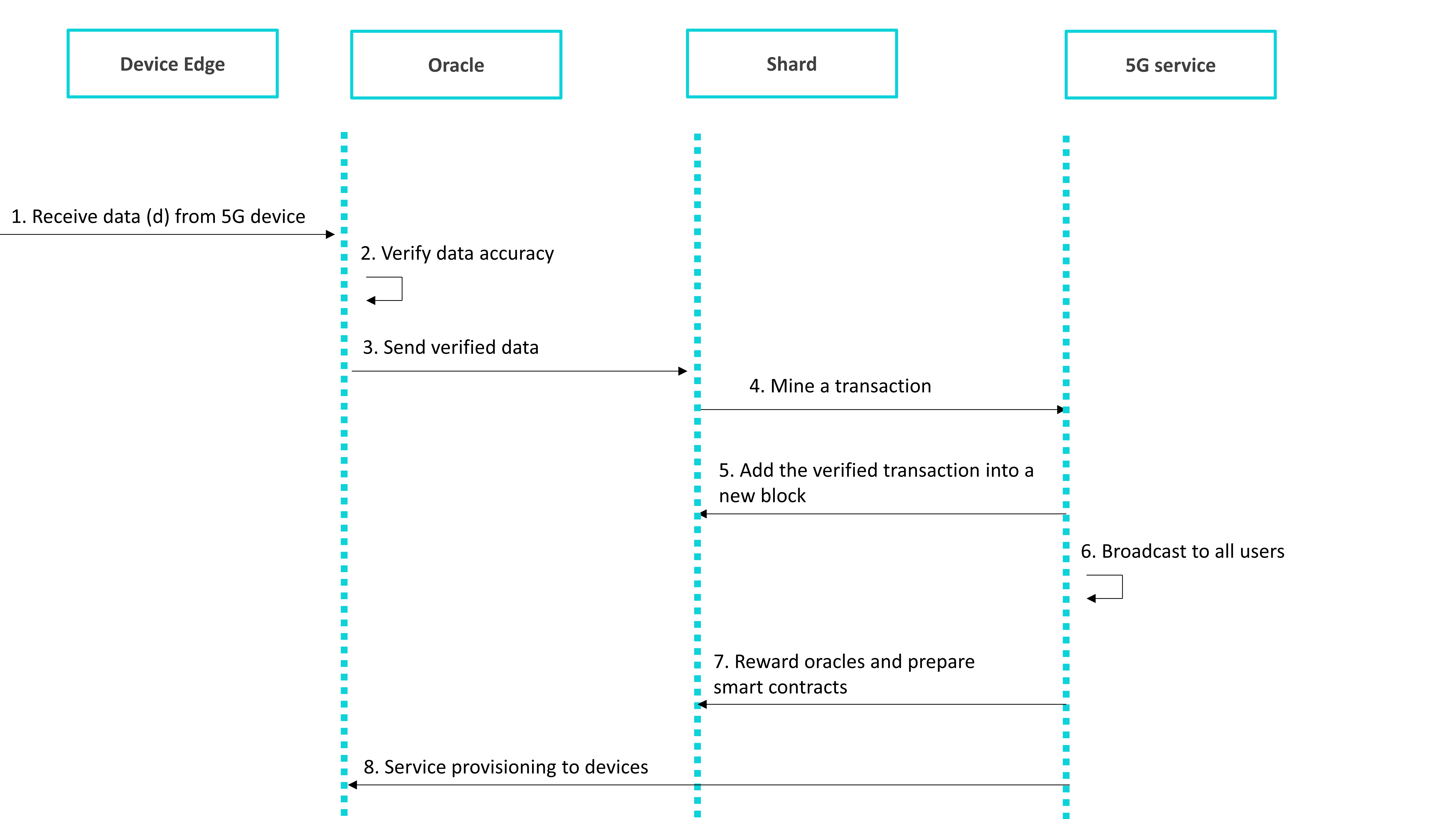}
	\caption{ A sequence diagram of utilizing oracles and shards with 5G services.}
	\label{fig:archi}
\end{figure}
Broadly speaking, nodes ($i.e.$, miners) are rewarded for transaction verification in which they took part. Conversely, those who provided false verification are penalized. The steps of reward/penalize are depicted in Fig.6. Let $r_i$ denote the reward amount that a node $i$ used to verify that a transaction partition is equal to $0$ or $1$. In the case of a valid verification, the reward is as follows: 
\begin{equation}
r_i=\frac{R(T_{i,j})}{Total} + \alpha_i
\end{equation}
where, $R(T_{i,j})$ is the master cluster reward, $Total$ refers to the total number of the cluster devices participating in the verification process, and $\alpha_i$ node deposit. A node's reward is equal to his participating share in the transaction verification process. For instance, if a cluster has a reward equal to 1000, and the number of participant nodes is equal to 100, the reward is distributed equally, and each node will receive 100 as a stake. In the case of false verification, the node is penalized, and the reward is deducted from its deposit.  

\section{Challenges and Future Research Directions}
Blockchain technology has the ability to improve 5G networks and to solve several issues ranging from security challenges to system management in a decentralized manner. Yet, various challenges need further investigation before the deployment of blockchain in 5G networks. This section highlights and discusses some of the open challenges that may set back the merging of blockchain with 5G networks.
\subsection{Scalability and Performance}
The scalability of blockchain is measured by the rate at which transactions are added to the chain, in other words, the number of transactions per second (TPS). Currently, popular blockchain platforms, such as Bitcoins and Ethereum, can reach up to 14 TPS, whereas some private blockchain can reach up to 3000 TPS. Transaction speed is one of the major concerns for adopting blockchain technology. High TPS is essential for 5G networks to meet the requirements of low latency and high communication rate. Straightforward integration of the current blockchain is unsuitable for 5G networks. Furthermore, blockchain has high throughputs because of the complexity of blockchain consensus. 5G networks require coordination between a large number of entities, services, and operations that cannot be achieved using blockchain. To solve these problems, further studies are required to improve the scalability of blockchain solutions to meet the requirement of 5G in a dynamic and heterogeneous manner and to involve a large number of transactions with low latency.
\subsection{Standardization and Regulations}

It is expected that various services and applications are going to be delivered by the 5G systems. However, standardizations are still challenging because of the interoperability, ubiquitously, and dynamicity of 5G communication systems and applications. On the other hand, blockchain just started gaining the attention of network operators. Consequently, to get a wide acceptance, network operators should work together to define a standardization of blockchain integration with 5G networks. Furthermore, blockchain smart contracts are de-standardized and deregulated, and unmodifiable, it is important to ensure the security of blockchain technology before its integration with 5G networks. Any falsified manipulation of blockchain smart contracts can severely damage the network and lead to serious consequences. For a large adoption of blockchain in the context of 5G networks, standardization, regulation, and governance must be enacted. 
\subsection{Resource Constraints}
The blockchain was designed for an internet scenario with powerful computers; it is computationally expensive and has significant overhead in terms of both bandwidth and storage capacity. These particular characteristics currently exclude the easy integration of the blockchain with 5G networks. Indeed, to take part in the blockchain nodes ($i.e.$, miners that can verify a transaction) need to have a high-computing power to run a blockchain consensus \cite{r16}. In some situations, 5G nodes may be already taking part in an operation to provide a service and might not be able to have enough resources to run also the blockchain, which may lead to network degradation, bottleneck, and SPF issues. Due to resource limitations, a lightweight framework that can dynamically optimize resource usage among several nodes is required for the 5G networks. Consequently, resource provisioning for restrained nodes should be further investigated. Besides, IoT devices are resource-constrained. Incorporating blockchain technology with the IoT nodes is challenging. The high computing power storage capacity and overheads have put under question the feasibility of the integration of blockchain within the 5G \cite{icc2}.  
\subsection{Interoperability}
Achieving seamless interoperability among blockchain and 5G networks is still challenging. Indeed, several well-known blockchain platforms offer different features. Nevertheless, it is unclear how blockchain technology can be merged with 5G networks \cite{icc}. Furthermore, 5G networks and beyond involve various new technologies such as NFV, SDN, MEC, Massive MIMO, and full duplex \cite{r25} \cite{r26} \cite{r24} \cite{icc1}. Each of which works differently. These are key challenges that need to be solved before the integration. Among the questions that are worth investigating are: 1) how can we deploy blockchain in 5G networks? Who will be responsible for ensuring the interoperability among 5G services, operations, and users? Is blockchain going to be deployed as an overlay network?
\subsection{Security}
Even if blockchain is considered as a secure technology that establishes trust in 5G networks, it still suffers from various security issues, such as 51 \% attacks that occur when an attacker takes positing of 51 \% of all network computing power. This type of attack led to falsification, SPF, and data corruption. If one attacker takes the position of 51 \% computing power, then he can control all the blockchain and thus all services and operations running in the blockchain. Furthermore, blockchain smart contracts can cause security issues due to poorly written code. Safe deployment of blockchain technology with a 5G network should take into consideration the issue inherent to the blockchain. Furthermore, data privacy has become a major concern especially in the context of blockchain and 5G. For example, data related to a credit card used to pay for a service provided by the 5G network can be stored in the blockchain permanently and cannot be deleted. By design, private data should not be stored in the chain but rather out of the chain and only have a pointer of that information. In this context, several researchers have suggested the use of on-the-chain and off-the-chain storage, on the chain stores will only have pointers, metadata, and hashes of the actual data stored off-the-chain in a secure database. 
\subsection{Infrastructure Costs}
It is expected that 5G networks are going to increase energy consumption because of the increased amount of equipment \cite{icc4} \cite{icc3}. With ongoing efforts for energy-efficient communications and networking, green communications are going to be harder to achieve. The blockchain solution will be a hurdler for 5G networks because of its computationally intensive consensus algorithms. For example, the bitcoin blockchain is estimated to consume at a peak of more electricity than 159 countries. Furthermore, the use of blockchain platforms and cloud infrastructures that host blockchain nodes comes at a cost. In some situations, every blockchain transaction has fees. For example, in the Ethereum blockchain, a transaction can cost a gas unit which refers to the number of computation efforts required to verify a transaction. Transaction costs are relative to the complexity of the consensus running. Consequently, designing an energy-efficient blockchain consensus algorithm is a big hurdle for blockchain integration in the 5G networks.
\section{Conclusion}
Blockchain technology was originally intended for cryptocurrency context, however, this technology has moved beyond its realm. Blockchain-based solutions have benefited from blockchain features to improve their process. In this context, blockchain was proposed by several researchers as a solution to 5G inherent challenges. Indeed, several studies have shown the benefit of using blockchain solutions to meet the requirements of 5G networks such as security, transparency, decentralization, and immutability. Blockchain integration with 5G networks is expected to improve new technologies with a secure design concept, limited access control, and trust. Furthermore, blockchain smart contract, lightweight consensus, sharding concept, and oracle blockchain will enable several new 5G business models to benefit from a high data rate, secure communication, and reliable services. In this chapter, we have reviewed some blockchain integration opportunities that can empower 5G network services and operations. Based on this overview we have proposed a scalable, secure, and lightweight blockchain architecture suitable with 5G requirements. Furthermore, we summarized some open challenges and provides some research directions that need further investigation for the safe deployment of blockchain in 5G networks.
\section*{acknowledgement}
\label{sec:acknowledgement}
The authors would like to thank the Natural Sciences and Engineering Research Council of Canada, as well as FEDER and GrandEst Region in France, for the financial support of this research.
\bibliographystyle{IEEEtran}
\bibliography{references}

\begin{thebibliography}{10}
\providecommand{\url}[1]{#1}
\csname url@samestyle\endcsname
\providecommand{\newblock}{\relax}
\providecommand{\bibinfo}[2]{#2}
\providecommand{\BIBentrySTDinterwordspacing}{\spaceskip=0pt\relax}
\providecommand{\BIBentryALTinterwordstretchfactor}{4}
\providecommand{\BIBentryALTinterwordspacing}{\spaceskip=\fontdimen2\font plus
\BIBentryALTinterwordstretchfactor\fontdimen3\font minus
  \fontdimen4\font\relax}
\providecommand{\BIBforeignlanguage}[2]{{%
\expandafter\ifx\csname l@#1\endcsname\relax
\typeout{** WARNING: IEEEtran.bst: No hyphenation pattern has been}%
\typeout{** loaded for the language `#1'. Using the pattern for}%
\typeout{** the default language instead.}%
\else
\language=\csname l@#1\endcsname
\fi
#2}}
\providecommand{\BIBdecl}{\relax}
\BIBdecl

\bibitem{s1}
A.~Alalewi, I.~Dayoub, and S.~Cherkaoui, ``On {5G}-{V2X} {Use} {Cases} and
  {Enabling} {Technologies}: {A} {Comprehensive} {Survey},'' \emph{IEEE
  Access}, vol.~9, pp. 107\,710--107\,737, 2021, conference Name: IEEE Access.

\bibitem{r3}
A.~{Gupta} and R.~K. {Jha}, ``A survey of 5g network: Architecture and emerging
  technologies,'' \emph{IEEE Access}, vol.~3, pp. 1206--1232, 2015.

\bibitem{r27}
Z.~Mlika and S.~Cherkaoui, ``Massive {Access} in {Beyond} {5G} {IoT} {Networks}
  with {NOMA}: {NP}-hardness, {Competitiveness} and {Learning},''
  \emph{arXiv:2002.07957 [cs, math]}, Feb. 2020, arXiv: 2002.07957.

\bibitem{r4}
X.~Li, P.~Jiang, T.~Chen, X.~Luo, and Q.~Wen, ``A survey on the security of
  blockchain systems,'' \emph{Future Generation Computer Systems}, vol. 107,
  pp. 841--853, 2020.

\bibitem{bb}
z.~Mlika and S.~Cherkaoui, ``Competitive {Algorithms} and {Reinforcement}
  {Learning} for {NOMA} in {IoT} {Networks},'' in \emph{{ICC} 2021 - {IEEE}
  {International} {Conference} on {Communications}}, Jun. 2021, pp. 1--6.

\bibitem{r5}
I.-C. Lin and T.-C. Liao, ``\BIBforeignlanguage{en}{A {Survey} of {Blockchain}
  {Security} {Issues} and {Challenges}},''
  \emph{\BIBforeignlanguage{en}{International Journal of Network Security}},
  vol.~19, no.~5, pp. 653--659, Sep. 2017.

\bibitem{r6}
A.~{Chaer}, K.~{Salah}, C.~{Lima}, P.~P. {Ray}, and T.~{Sheltami}, ``Blockchain
  for 5g: Opportunities and challenges,'' in \emph{2019 IEEE Globecom Workshops
  (GC Wkshps)}, 2019, pp. 1--6.

\bibitem{r23}
Z.~A.~E. {Houda}, A.~{Hafid}, and L.~{Khoukhi}, ``Blockchain meets ami: Towards
  secure advanced metering infrastructures,'' in \emph{ICC 2020 - 2020 IEEE
  International Conference on Communications (ICC)}, 2020, pp. 1--6.

\bibitem{r7}
D.~C. Nguyen, P.~N. Pathirana, M.~Ding, and A.~Seneviratne, ``Blockchain for
  {5G} and {Beyond} {Networks}: {A} {State} of the {Art} {Survey},''
  \emph{arXiv:1912.05062 [cs, eess, math]}, Dec. 2019, arXiv: 1912.05062.

\bibitem{r22}
Z.~{Abou El Houda}, A.~S. {Hafid}, and L.~{Khoukhi}, ``Cochain-sc: An intra-
  and inter-domain ddos mitigation scheme based on blockchain using sdn and
  smart contract,'' \emph{IEEE Access}, vol.~7, pp. 98\,893--98\,907, 2019.

\bibitem{r10}
H.~{Moudoud}, S.~{Cherkaoui}, and L.~{Khoukhi}, ``An iot blockchain
  architecture using oracles and smart contracts: the use-case of a food supply
  chain,'' in \emph{2019 IEEE 30th Annual International Symposium on Personal,
  Indoor and Mobile Radio Communications (PIMRC)}, 2019, pp. 1--6.

\bibitem{r11}
B.~Liu, X.~L. Yu, S.~Chen, X.~Xu, and L.~Zhu, ``Blockchain {Based} {Data}
  {Integrity} {Service} {Framework} for {IoT} {Data},'' \emph{2017 IEEE
  International Conference on Web Services (ICWS)}, 2017.

\bibitem{r21}
Z.~{Abou El Houda}, L.~{Khoukhi}, and A.~{Senhaji Hafid}, ``Bringing
  intelligence to software defined networks: Mitigating ddos attacks,''
  \emph{IEEE Transactions on Network and Service Management}, vol.~17, no.~4,
  pp. 2523--2535, 2020.

\bibitem{r12}
X.~Li, P.~Jiang, T.~Chen, X.~Luo, and Q.~Wen, ``\BIBforeignlanguage{en}{A
  survey on the security of blockchain systems},''
  \emph{\BIBforeignlanguage{en}{Future Generation Computer Systems}}, vol. 107,
  pp. 841--853, Jun. 2020.

\bibitem{r15}
X.~Xu, I.~Weber, M.~Staples, L.~Zhu, J.~Bosch, L.~Bass, C.~Pautasso, and
  P.~Rimba, ``A {Taxonomy} of {Blockchain}-{Based} {Systems} for {Architecture}
  {Design},'' in \emph{2017 {IEEE} {International} {Conference} on {Software}
  {Architecture} ({ICSA})}, Apr. 2017, pp. 243--252.

\bibitem{r13}
N.~M. Kumar and P.~K. Mallick, ``Blockchain technology for security issues and
  challenges in {IoT},'' \emph{Procedia Computer Science}, vol. 132, pp.
  1815--1823, Jan. 2018.

\bibitem{r16}
Z.~{Abou El Houda}, A.~S. {Hafid}, and L.~{Khoukhi}, ``Cochain-sc: An intra-
  and inter-domain ddos mitigation scheme based on blockchain using sdn and
  smart contract,'' \emph{IEEE Access}, vol.~7, pp. 98\,893--98\,907, 2019.

\bibitem{r14}
L.~S. Sankar, M.~Sindhu, and M.~Sethumadhavan, ``\BIBforeignlanguage{en}{Survey
  of consensus protocols on blockchain applications},'' in
  \emph{\BIBforeignlanguage{en}{2017 4th {International} {Conference} on
  {Advanced} {Computing} and {Communication} {Systems} ({ICACCS})}}.\hskip 1em
  plus 0.5em minus 0.4em\relax Coimbatore, India: IEEE, Jan. 2017, pp. 1--5.

\bibitem{r17}
Z.~A. {El Houda}, L.~{Khoukhi}, and A.~{Hafid}, ``Chainsecure - a scalable and
  proactive solution for protecting blockchain applications using sdn,'' in
  \emph{2018 IEEE Global Communications Conference (GLOBECOM)}, 2018, pp. 1--6.

\bibitem{icc6}
Z.~Mlika and S.~Cherkaoui, ``Massive {IoT} {Access} {With} {NOMA} in {5G}
  {Networks} and {Beyond} {Using} {Online} {Competitiveness} and {Learning},''
  \emph{IEEE Internet of Things Journal}, vol.~8, no.~17, pp. 13\,624--13\,639,
  Sep. 2021.

\bibitem{r19}
H.~{Moudoud}, S.~{Cherkaoui}, and L.~{Khoukhi}, ``Towards a scalable and
  trustworthy blockchain: Iot use case,'' in \emph{ICC 2021 - 2021 IEEE
  International Conference on Communications (ICC)}, 2021, pp. 1--6.

\bibitem{r18}
H.~{Moudoud}, L.~{Khoukhi}, and S.~{Cherkaoui}, ``Prediction and detection of
  fdia and ddos attacks in 5g enabled iot,'' \emph{IEEE Network}, pp. 1--8,
  2020.

\bibitem{z1}
Z.~Mlika and S.~Cherkaoui, ``Massive {Access} in {Beyond} {5G} {IoT} {Networks}
  with {NOMA}: {NP}-hardness, {Competitiveness} and {Learning},''
  \emph{arXiv:2002.07957 [cs, math]}, Feb. 2020, arXiv: 2002.07957.

\bibitem{icc2}
A.~Rachedi, M.~H. Rehmani, S.~Cherkaoui, and J.~J. P.~C. Rodrigues, ``{IEEE}
  {Access} {Special} {Section} {Editorial}: {The} {Plethora} of {Research} in
  {Internet} of {Things} ({IoT}),'' \emph{IEEE Access}, vol.~4, pp. 9575--9579,
  2016.

\bibitem{icc}
A.~Abouaomar, M.~Elmachkour, A.~Kobbane, H.~Tembine, and M.~Ayaida,
  ``Users-{Fogs} association within a cache context in {5G}
  networks:{Coalition} game model,'' in \emph{2018 {IEEE} {Symposium} on
  {Computers} and {Communications} ({ISCC})}, Jun. 2018, pp. 00\,014--00\,019,
  iSSN: 1530-1346.

\bibitem{r25}
Z.~A.~E. {Houda}, A.~{Hafid}, and L.~{Khoukhi}, ``Blockchain-based reverse
  auction for v2v charging in smart grid environment,'' in \emph{ICC 2021 -
  2021 IEEE International Conference on Communications (ICC)}, 2021, pp. 1--6.

\bibitem{r26}
A.~{Filali}, Z.~{Mlika}, S.~{Cherkaoui}, and A.~{Kobbane}, ``Preemptive sdn
  load balancing with machine learning for delay sensitive applications,''
  \emph{IEEE Transactions on Vehicular Technology}, vol.~69, no.~12, pp.
  15\,947--15\,963, 2020.

\bibitem{r24}
Z.~A. {El Houda}, A.~{Hafid}, and L.~{Khoukhi}, ``Co-iot: A collaborative ddos
  mitigation scheme in iot environment based on blockchain using sdn,'' in
  \emph{2019 IEEE Global Communications Conference (GLOBECOM)}, 2019, pp. 1--6.

\bibitem{icc1}
E.~D. Ngangue~Ndih and S.~Cherkaoui, ``On {Enhancing} {Technology}
  {Coexistence} in the {IoT} {Era}: {ZigBee} and 802.11 {Case},'' \emph{IEEE
  Access}, vol.~4, pp. 1835--1844, 2016.

\bibitem{icc4}
E.~D.~N. Ndih and S.~Cherkaoui, ``\BIBforeignlanguage{en}{Chapter 17 -
  {Simulation} methods, techniques and tools of computer systems and
  networks},'' in \emph{\BIBforeignlanguage{en}{Modeling and {Simulation} of
  {Computer} {Networks} and {Systems}}}, M.~S. Obaidat, P.~Nicopolitidis, and
  F.~Zarai, Eds.\hskip 1em plus 0.5em minus 0.4em\relax Boston: Morgan
  Kaufmann, Jan. 2015, pp. 485--504.

\bibitem{icc3}
E.~D. Ngangue~Ndih, S.~Cherkaoui, and I.~Dayoub, ``Analytic {Modeling} of the
  {Coexistence} of {IEEE} 802.15.4 and {IEEE} 802.11 in {Saturation}
  {Conditions},'' \emph{IEEE Communications Letters}, vol.~19, no.~11, pp.
  1981--1984, Nov. 2015.

\end{thebibliography}

\end{document}